\documentclass[12pt]{article}   
\usepackage{psfig,times,fancyheadings}  % 17 dec. 2001: pbs avec times
\begin{document}
\thispagestyle{empty}

 \lhead[\fancyplain{}{\sl }]{\fancyplain{}{\sl }}
 \rhead[\fancyplain{}{\sl }]{\fancyplain{}{\sl }}

%%%%%%%% Pour changer les valeurs par defaut pour taille figure,
%%%%%%%% sinon au-dela d'une hauteur de 134 mm = 70% on est rejete a la fin
 \renewcommand{\topfraction}{.99}      
 \renewcommand{\bottomfraction}{.99} 
 \renewcommand{\textfraction}{.0}

%%%%% Definitions

\newcommand{\nc}{\newcommand}

\nc{\qI}[1]{\section{{#1}}}
\nc{\qA}[1]{\subsection{{#1}}}
\nc{\qun}[1]{\subsubsection{{#1}}}
\nc{\qa}[1]{\paragraph{{#1}}}

            % Enumerations
\def\qbu{\hfill \par \hskip 6mm $ \bullet $ \hskip 2mm}
\def\qee#1{\hfill \par \hskip 6mm #1 \hskip 2 mm}

\nc{\qfoot}[1]{\footnote{{#1}}}
\def\qL{\hfill \break}
\def\qpar{\vskip 2mm plus 0.2mm minus 0.2mm}
\def\qtvi{\vrule height 2pt depth 5pt width 0pt}
\def\qth{\vrule height 12pt depth 0pt width 0pt}
\def\qtb{\vrule height 0pt depth 5pt width 0pt}
\def\tvi{\vrule height 12pt depth 5pt width 0pt}

\def\qparr{ \vskip 1.0mm plus 0.2mm minus 0.2mm \hangindent=10mm
\hangafter=1}

                % Decale UN paragraphe
                % Attention! La double accolade est vitale, sinon tout le
                % est decale (cf TEX p.199)
                % On peut aller a la ligne avec \qL=\hfill \break
                % Par contre ne supporte pas les lignes blanches
\def\qdec#1{\par {\leftskip=2cm {#1} \par}}

   %% Defs specifiques
\def\qdpt{\partial_t}
\def\qdpx{\partial_x}
\def\qddpt{\partial^{2}_{t^2}}
\def\qddpx{\partial^{2}_{x^2}}
\def\qn#1{\eqno \hbox{(#1)}}
\def\qds{\displaystyle}
\def\qw{\widetilde}
\def\qmax{\mathop{\rm Max}}   % Petit livre Tex (p.167)
\def\qmin{\mathop{\rm Min}}   % Petit livre Tex (p.167)

%%%%% End of definitions

\def\qci#1{\parindent=0mm \par \small \parshape=1 1cm 15cm  #1 \par
               \normalsize}

\null

\centerline{\bf \Large Does the price multiplier effect also hold for stocks?}
\vskip 0.5cm
 \centerline{\bf \Large }

\vskip 1cm
\centerline{\bf Sergei Maslov $ ^1 $ and Bertrand M. Roehner $ ^2 $ }
\vskip 4mm         
%\centerline{\bf Institute for Theoretical and High Energy Physics}
%\centerline{\bf University Paris 7 }

\vskip 2cm

{\bf Abstract}\quad  The price multiplier effect 
provides precious insight
into the behavior of investors during episodes of speculative trading.
It tells us that the higher the price of an asset is (within a set of
similar assets) the more its price is likely to 
increase during the upgoing
phase of a speculative price peak. In short, instead of being 
risk averse, as is often assumed, investors rather seem to be
``risk prone''. While
this effect is known to  hold for several sorts of assets, it has
not yet been
possible to test it for stocks because
the price of one share has no intrinsic significance which
means that one cannot say that a stock $ A $ is more expensive
than a stock $ B $ on the basis of its price. 
In this paper we
show that the price-dividend ratio gives a good basis 
for assessing the price of stocks in an intrinsic way. 
When this alternative measure is used instead,
it turns out that the price 
multiplier effect also holds for stocks, at least if one concentrates
on samples of companies which are sufficiently homogeneous. 

\vskip 1cm

\centerline{19 May 2003}

\vskip 8mm
%\centerline{\it Preliminary version}
\vskip 8mm
Keywords: stock prices, speculation, real estate, price-dividend ratio

\vskip 2cm

1: Department of Physics,  Brookhaven National Laboratory, Upton, 
New York 11973; maslov@bnl.gov
\vskip 4mm
2: Permanent affiliation: Institute for Theoretical and High Energy Physics,
University of Paris. 
\qL
\phantom{1: }Postal address to which correspondence should be sent:\qL
\phantom{1: }Bertrand Roehner, LPTHE, University Paris 7, 2 place Jussieu, 
75005 Paris, France.
\qL
\phantom{1: }E-mail: ROEHNER@LPTHE.JUSSIEU.FR
\qL
\phantom{1: }FAX: 33 1 44 27 79 90
\qL
\phantom{1: }{\it This research was done during a stay of this 
author
as a visiting researcher at Brookhaven Lab in April-May 2003.}
\vfill \eject

\qI{Introduction}

The price multiplier effect gives us an insight into the behavior
of investors during a speculative price peak. It tells us
that the higher the price of an asset, the higher the amplitude
of the price peak by which we understand the ratio $ A=p_2/p_1 $ of the
peak price $ p_2 $ to the initial price $ p_1 $. In short,
highly priced items seem to attract speculative capital in the
same way as a magnet would attract bits of iron. 
In previous works
(see [1]) this effect was shown to 
hold (during speculative episodes) for 
items as diverse as real estate, collector's stamps,
antiquarian books or rare coins. What do these items have
in common? Firstly, all of them happen to experience price peaks 
of substantial amplitude (say $ A>2 $ ); a second common feature
is the fact that an investor can select one item 
among similar ones at almost no cost. For instance,
an investor in stamps can choose whether to buy 20 stamps
valued approximately ten euros each or instead 
to buy a single stamp valued
at 200 euros. The only cost involved will
be the time it takes to make the selection (the transaction cost
may be slightly different also but we will neglect this effect).
In contrast,
for commodities a very different situation prevails: for instance
it is a very different matter to have a cargo of oil in the
Persian Gulf valued at \$20/barrel, or one in the Gulf of
Mexico valued at \$25/barrel for in this case there is not only a
difference in price but also a difference in location; to 
substitute one cargo for another would involve substantial
transportation costs. 

A natural question is whether or not the price multiplier effect
also holds for stocks. Answering that question is the purpose of
this paper; it is not an easy problem for at least three reasons.
\qbu First one has to define how
to compare the price of two stocks. Whereas it makes sense to say
that a given house in San Francisco costs twice as much as 
another of same size in Oklahoma City, it is meaningless to say 
that the price of an IBM stock is twice the price of a
CISCO stock for, as we know, through stock splits stock prices
can be reduced 
by an arbitrary factor almost overnight.
To sum up, our very
first task will be to define the price of stocks in an intrinsic
way so that they can
be compared one with another in a meaningful way.
\qbu The second difficulty 
has to do with the diversity which is inherent to stock markets. 
In order to explain this point let us again consider the
example of the two houses in San Francisco and Oklahoma City;
despite their disparity
the prices of the two assets are determined by the same set of rules, 
namely by the rules which govern residential
property markets. On the contrary, two companies such as Yahoo  
(an Internet company) 
and Burlington Northern Santa Fe (a railroad company) belong 
to two very different worlds and it is unlikely that their stock prices
will be determined by similar rules. Naturally, to some extent it is
possible to remedy to this difficulty by selecting stocks
from a specified sample. 
For instance one may consider stocks belonging to the Dow Jones
Transportation index that all belong to the transportation industry.
But even for such
a restricted sample there will be companies as different 
as Delta Airlines and the above mentioned Burlington railroad 
company. 
\qbu The third difficulty stems from the well-know fact that
stock prices have a much higher volatility than 
housing prices. For instance the stock price of Wall-Mart 
(a retail trade company) increased from \$ 0.12  in January 
1980 to \$ 61 dollar in March 2002; once deflated this represents a 
multiplication by a factor 239. The case of Wal-Mart is by no
means unique for as we know
during the 1980-2000 time interval, not a few
companies experienced huge price peaks, especially in the
telecommunication and information technology sectors.
Needless to say, price 
increases of such an amplitude are completely unheard of in
the housing market; this illustrates the fact
that stock prices are much more volatile than housing prices,
a feature which is directly connected with the fact that transaction
costs and transaction times in housing markets are larger
than in stock markets by several orders of magnitude.
Needless to say, the high volatility of stock prices will
add a substantial amount of noise and make it more
difficult to test the price multiplier effect.
\qpar

In the second section, starting from
the known case of real estate markets, 
we show how the price
multiplier effect can be extended to stocks.
In the third section we explain why the price-dividend
ratio can be seen as a convenient yardstick for 
assessing the respective value of stocks.
Then in the fourth section we present the results
of a several statistical tests of the price multiplier
effect in stocks.

\qI{From real estate to stocks}
First of all, in order for the reader to be able to
develop an intuitive understanding of 
the price multiplier effect and its implications,
we briefly illustrate it by way of an example. We consider
housing prices in the West of the United States
over the 1995-2002 time interval. Fig.1 shows that there is a 
significant correlation between initial prices in 12 major cities
at the beginning of the price peak (that is to say in 1995) and
the amplitudes of the price peaks; the correlation is equal to
0.75 with a confidence interval (at probability level 0.95)
extending from 0.30 to 0.92. For California alone (6 cities)
the correlation is equal to 0.71. 
%%-----------------------------------------------
%%%% Fig.1
  \begin{figure}[tb]
    \centerline{\psfig{width=17cm,figure=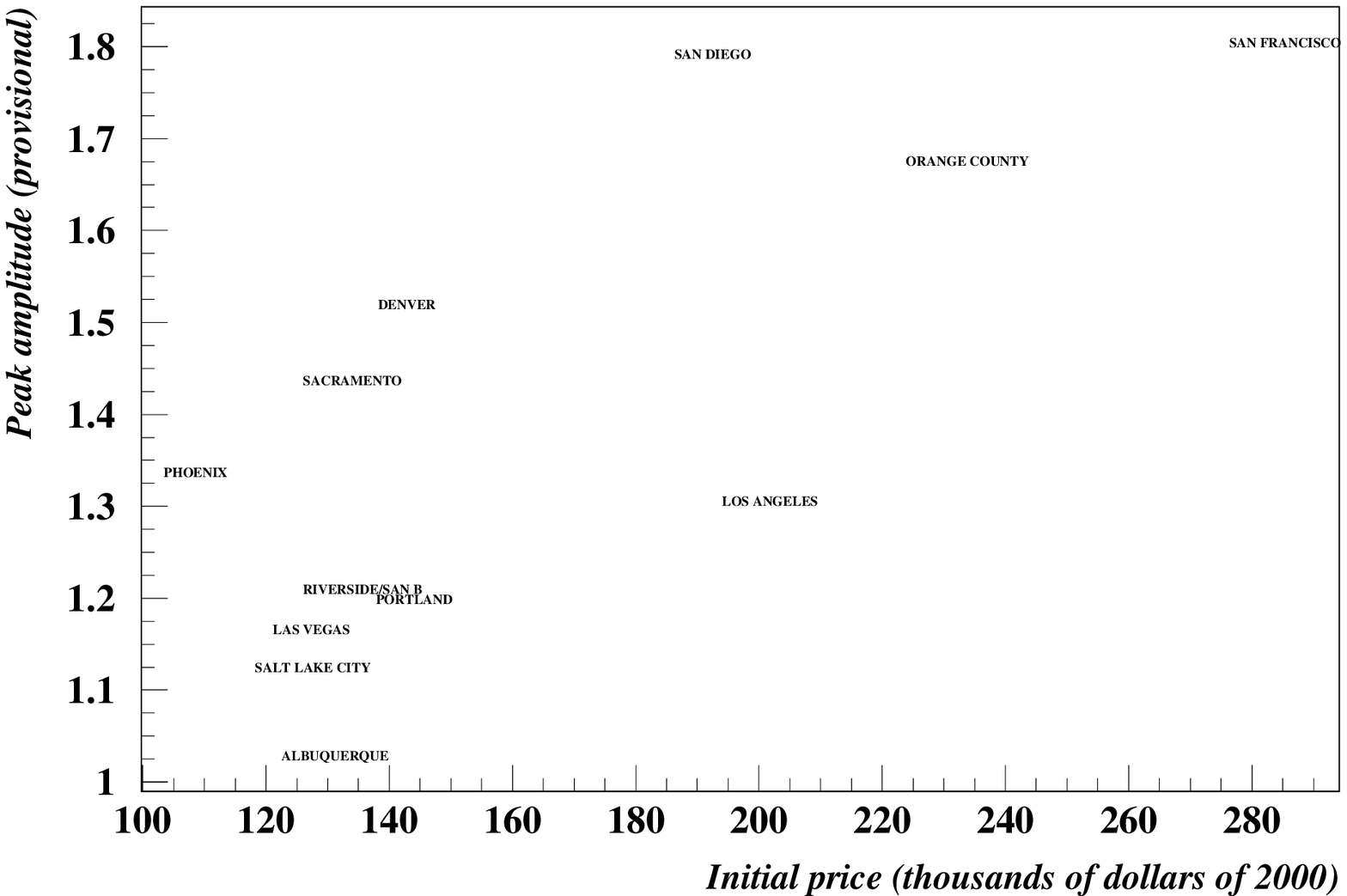}}
    {\bf Fig.1: Correlation between initial housing prices
and peak amplitudes, 1995-2002.} 
{\small The correlation is equal to 0.75.}
{\small \it Source: Website of the California Association of
Realtors}.
 \end{figure}
%% --------------------------------------------------
It can be noted that these
correlation levels are substantially higher than during
the previous peak of 1984-1990; for California the correlation was
then equal to only 0.60 (with a price increase of about 60 percent in
San Francisco);
moreover for the western cities located outside of
California, prices showed no substantial increases (less than 10 percent
over 6 years).
\qL
As an empirical rule, the level of the correlation
in the price multiplier effect provides a measure of how widespread
speculative trading is. If only we could  know the proportion of
transactions made by investors (as opposed to users),
we would be able to check that rule directly; unfortunately 
such data do not seem to be available either for California
or for any other housing market. The best possible check consists
in observing that usually the higher the price peaks, the higher
is also the correlation in the price multiplier effect. 
\qpar

Regarding Fig.1
the reader may object that in 2002 prices in the West were still 
increasing and that the amplitudes shown in Fig.1 represent only
a proxy of the actual peak amplitudes. This is true; in fact, it
is quite by purpose that we choose to present that case. It shows
that we do not
need to wait until all prices have peaked in order to test 
for the price multiplier effect. The obvious advantage of such
a procedure is that it 
provides an assessment of the level of speculative
trading at any moment during the upgoing phase. Naturally,
in the first quarters of 1995 price increases were still small as was
also the correlation of the price multiplier effect; but as
speculative trading built up and gained momentum
both price increases
and correlation became more substantial.  
\qpar

At the beginning of the paper we pointed out that the 
major obstacle for extending the price
multiplier effect to stocks is the fact that stock
prices have no intrinsic meaning. Can we reformulate
the price multiplier effect in the case of housing markets
by substituting another variable to initial housing prices?
The answer is yes: as will be seen,
prices can  be replaced by price-rent ratios, the latter being defined
as the the ratio of the price of the house (or the apartment) to
the annual rent paid by the tenant. This substitution is justified 
by the two following observations.
\qbu Prices are correlated with  price-rent ratios. This can
be illustrated by resorting again to the case of the West
in the U.S.
For 18 cities the correlation between prices and price-rent
ratios was equal to 0.66 in May 2001 and to 0.88 in May 2002
(more details can be found in [2]).
\qbu Fig.2 shows that during a 
speculative price peak, price and price-rent ratio move in
parallel. Thus, the evolution in the course of time
can be described either through price changes or through
variations in the price-rent ratio. 
In the legend of Fig.2 we used the expression 
price earnings ratio in order to underline the parallel
with the price-earnings ratio of stocks. 
The fact that the price-rent ratio 
increases in the ascending phase shows that rents either 
increase slower than prices or even decrease. Direct evidence
of a decrease can be found in New York City during the
price peak of 2001-2002: between mid-2001 and the end of 
2002, rents on average decreased by 15 percent while 
at the same time prices
increased by about 20 percent
(New York Times May 23, 2002, p.B9).
\qpar
%%-----------------------------------------------
%%%% Fig.2
  \begin{figure}[tb]
    \centerline{\psfig{width=17cm,figure=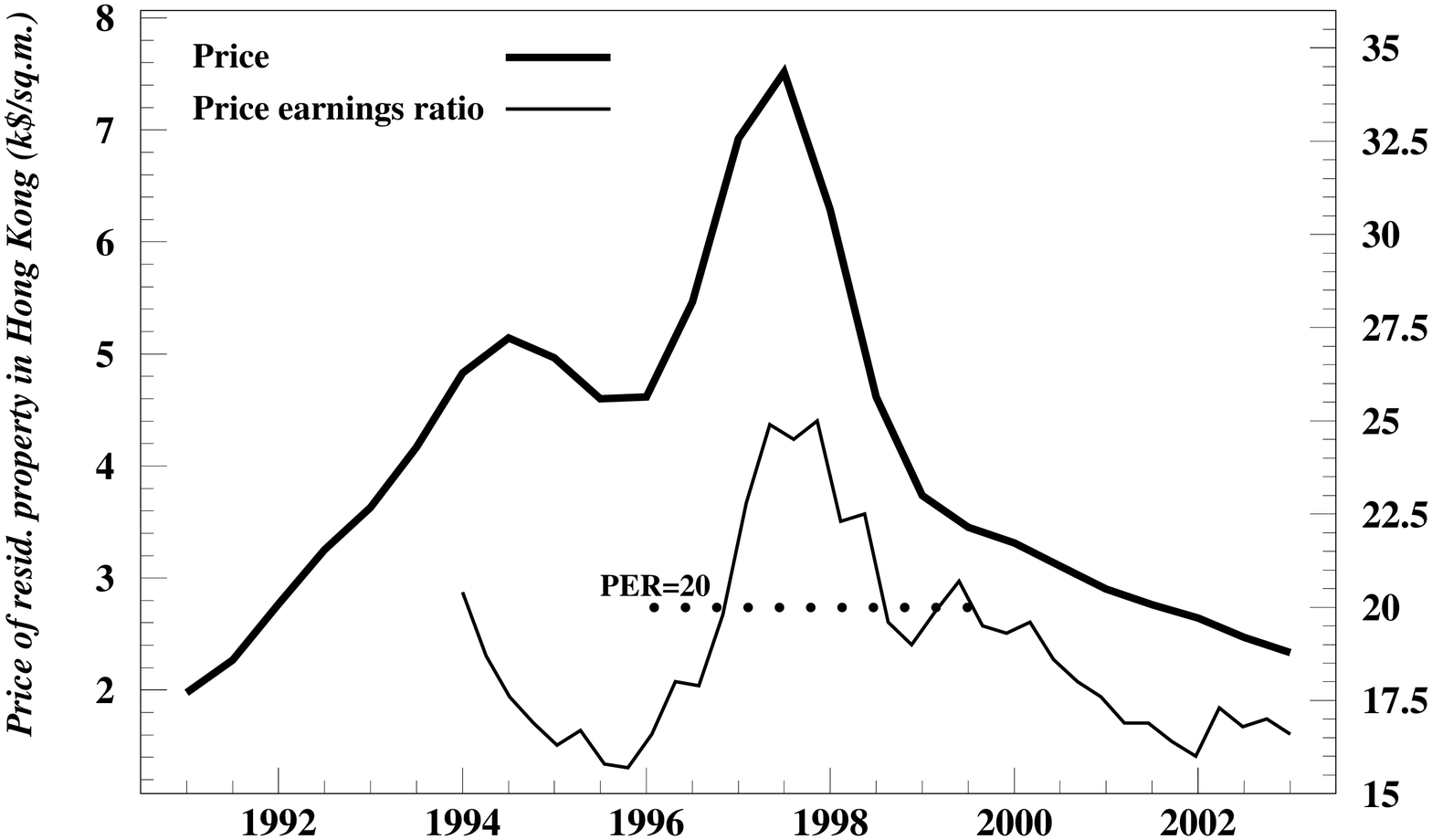}}
    {\bf Fig.2: Price and price-rent ratio in Hong Kong during
the speculative episode of 1992-2002} 
{\small The prices are expressed in $ 10^3 $ US\$ per 
square meter.
When seen from the owner's (instead of tenant's) perspective
the price-rent ratio should rather be called
a price earnings ratio; this is why we used this expression
in the graphic; it has the additional advantage of establishing
a clear link with the price-earnings ratio of stocks.}
{\small \it Source: Website of the Rating and Valuation 
Department of the Hong Kong Government}.
 \end{figure}
%% --------------------------------------------------

In short, at least during speculative episodes, prices
and price-rent ratios are strongly connected both temporally
and spatially; the great advantage of this observation comes
from the fact that, in contrast to prices, price-rent ratios
can be defined in an intrinsic way also for stocks. This is the 
point to which we come now.

\qI{Defining a yardstick for the price of stocks}

The yardstick we wish to define should be able
to tell us whether a stock $ A $ is cheaper or
more expensive than a stock $ B $, but that yardstick only
needs to apply to periods of speculative price peaks. 
Fig.3 a,b,c,d shows that during such episodes
prices and  price-dividend ratios move in parallel ways.

%%-----------------------------------------------
%%%% Fig.3
  \begin{figure}[page]
    \centerline{\psfig{width=15cm,figure=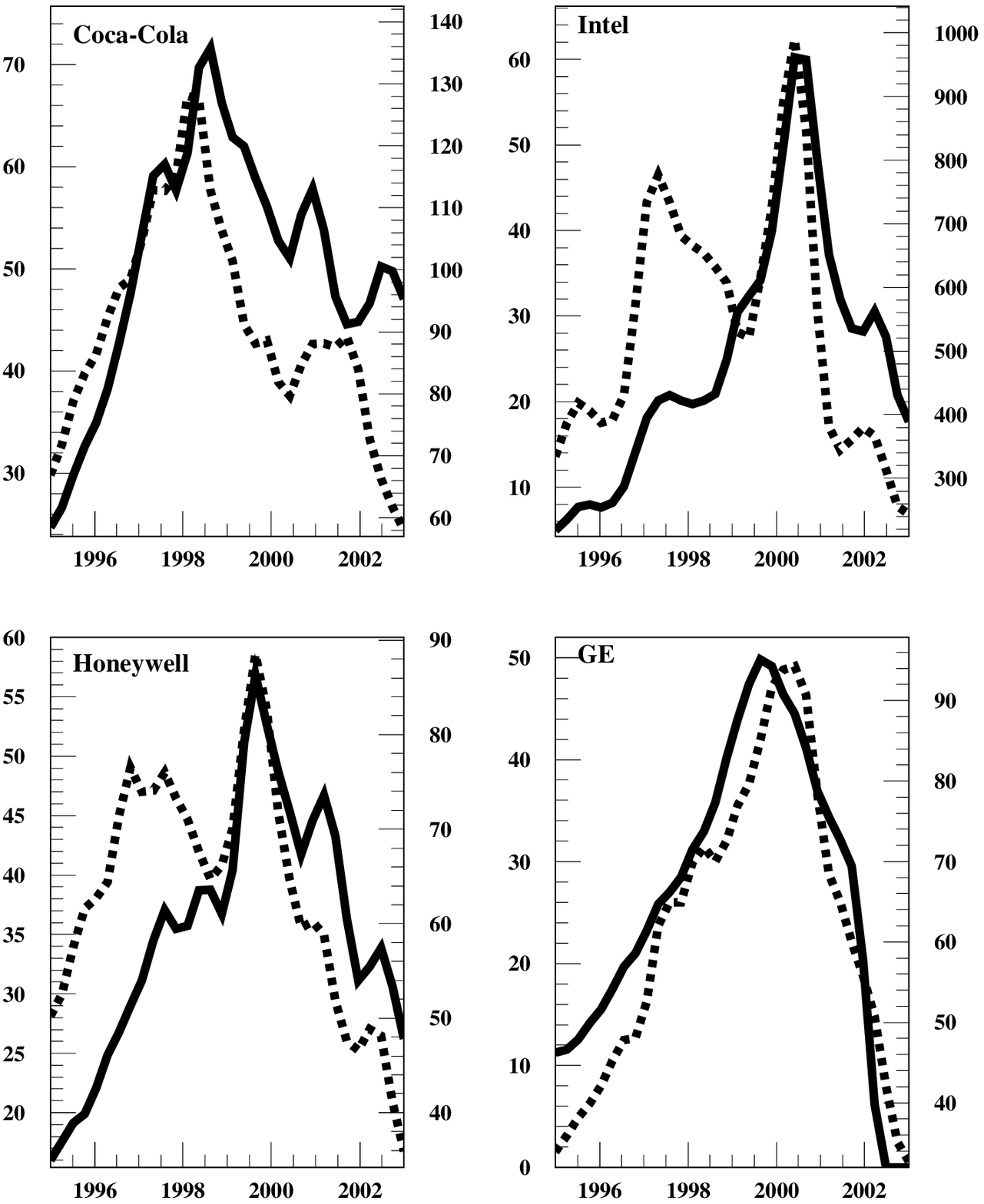}}
    {\bf Fig.3 a,b,c,d: Price versus price-dividend ratio for four
stocks.} 
{\small In the vicinity of the price peaks that occurred around
2000 the figure shows the respective evolution of the price 
(solid line and left-hand scale in dollars) 
and of the price-dividend ratio 
(broken line and right-hand side scale). The fact that the respective
curves are fairly parallel shows that the level of the dividends
remained
fairly constant as prices experienced a sharp increase or
decrease. Note the high level of the price-dividend ratio for
Intel Corporation, which is typical of the small dividends distributed
by the companies in the information technology.
GE stands for General Electric.}
{\small \it Source: http://finance.yahoo.com}

 \end{figure}
%% --------------------------------------------------

The definition of the price-dividend ratio used in these 
figures is the following:
$$ \hbox{PDR}(t) = { \hbox{Stock price at time}\ t \over
   \hbox{Dividends paid in the year preceding}\ t } $$

The price-dividend ratio is very similar to the price earnings
ratio but it has two clear advantages over the later (i) For a
stock holder it is the quarterly dividend 
(rather than the earnings per share) which is 
is the analogue of the monthly rent received by 
an apartment owner from his (or her) tenant.
(ii) Depending on the strategical accounting choices made by the
direction of a company the profit in a given quarter can
be reduced or inflated; for instance
writing off  a major acquisition over
four years instead of ten years would substantially lower the
earnings. Naturally, the level of dividend payments are also 
partly affected by the strategy of the company, but at least
the dividends represent real money and not just a line
in the company's accounting books. As the saying goes 
``Paper profits are one thing, cash is another.'' 
In short, relying on
dividends rather than on earnings seems to be a prudent and
conservative approach.
\qpar

On average
in normal times companies distribute about 60 percent of
their profit in the form of dividends, but for a company which wants
to invest aggressively, that proportion may be much smaller. 
It can be noted that on the basis of 60 percent of the earnings 
being distributed to stockholders, the price-dividend ratio should
be larger than the price earning ratio by a factor $ 1/0.60 = 1.70 $.
If, as is commonly admitted, the long term average of the price-earnings ratio over
the last century is about 14, the long term average of the PDR is
around $ 14\times 1.70=23 $. 
\qpar

To sum up, the price-dividend ratio fulfills all the 
requirements that one may wish for a yardstick of stock prices.
(i) During speculative episodes, PDR and stock price move in parallel.
(ii) Seen from the side of the
investor, the PDR is the exact analogue of the ratio of housing price
to annual rent; it tells the investor in how many years its purchase
(whether it is a house or a stock) will be covered by the flow of
rents or dividends. In short, it provides a measure of how expensive
the asset is with respect to future capital flow. 
(iii) The PDR (in contrast to the price-earnings ratio) can be measured in an 
objective and non-ambiguous way; furthermore long time series of
dividend data are
available in stock markets factbooks.
\qL
The question in the title of this paper
about whether or not the price multiplier effect holds for stocks
can therefore be rephrased in the following way: is there
a relationship between initial PDR levels and amplitudes of
price peaks?

\qI{Testing the price multiplier rule}

\qA{Setting up the experiment}
Is there a significant correlation between levels of the
price-dividend ratio at the start of a price peak and peak 
amplitudes, this is the question we wish to investigate.
However,
there are about 3,000 companies traded on the New York Stock Exchange
and 
the first question  we must address is how to select 
the sample(s) on which the price multiplier effect will be tested. 
\qL
There are several requirements.  
First, the selection must follow an objective criterion,
for example we may consider {\it all} the companies of the Dow Jones (DJ)
Industrial Index, or the first 30 companies of the Standard and Poor's
index arranged in alphabetical order.
In order to limit the diversity of the sample
we begin with the first choice. Besides the companies comprised
in the Dow Jones Industrial Index we will also investigate 
other fairly homogeneous sets of companies such as those
in the DJ Transportation index and the DJ Utility index.
\qL
The second question concerns the initial year, 
in other words when did the
speculative price peak begin? A possible criterion would be to consider
the largest time interval characterized by continuously
positive annual price variations
of stock indices. Such a criterion would put the beginning of
the price peak around 1980. However, since the decade before 1980 was
a bear market 
marked by substantial price falls, one may argue that during the early
1980s the market was merely catching up.
Whatever the validity of such a reasoning, it would anyway be difficult
to say with any degree of certainty whether the price peak has begun
in 1985, 1990 or 1995. This is why we will investigate all these dates
in the hope that the comparison of the different results may
give a better insight. 
\qpar

{\bf Remark} Special dividends may be
paid to stockholders on particular occasions, for instance
when the company sells a subsidiary. 
In some cases these special dividends are very large compared to
normal quarterly dividends. For instance, in May 1998 General
Motors paid a \$14.33 special dividend which represented
7 times the annual dividend paid in previous or later years;
other examples are Eastman Kodak which in January 1994 paid a special
dividend equal to 6 times the annual dividend or At\&T which in
October 2001 paid a special dividend equal to 11 times the annual
dividend. Clearly special dividends of such an order of magnitude
tend to disrupt the normal pattern of quarterly dividends over periods
of time which are of the order of one decade; 
thus, in order to avoid any bias we discarded the companies
and periods of time during which huge special dividends were 
distributed.

\qA{Results}
The relationship between price-dividend ratios and 
peak amplitudes 
is summarized in Fig.4a,b,c,d for the four initial years
1980, 1985, 1990, 1995. 
%%-----------------------------------------------
%%%% Fig.4
  \begin{figure}[page]
    \centerline{\psfig{width=13cm,figure=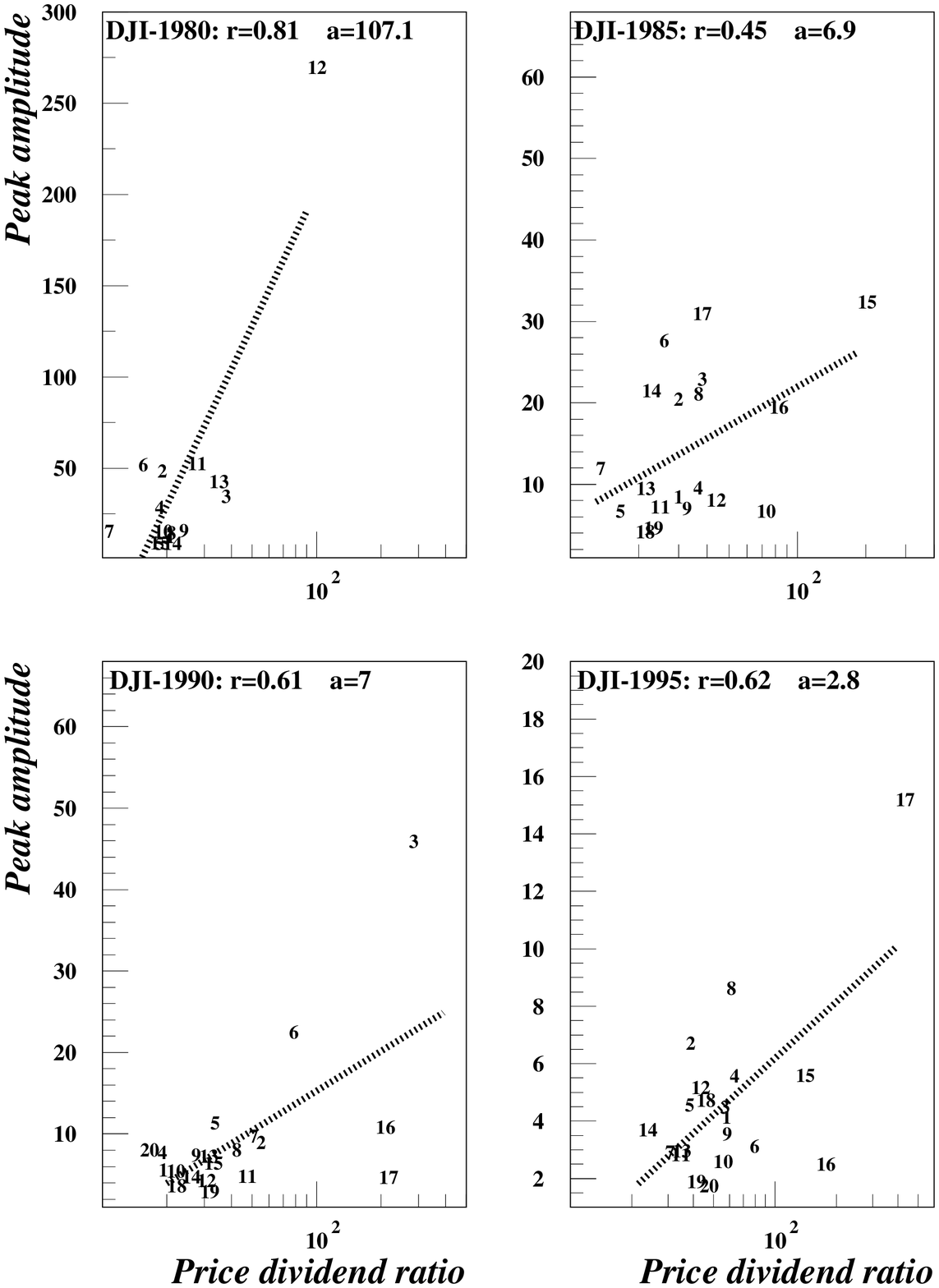}}
    {\bf Fig.4 a,b,c,d: Correlation between initial price-dividend
ratios and peak amplitudes.} 
{\small Horizontal axis: price-dividend ratio (logarithmic scale);
vertical scale: amplitude of the price peaks. 
The stocks are those comprised in the DJ Industrial index 
and for which price and dividend data were available on the Yahoo
website. In all cases there is a correlation which is significant at
probability level 0.95; however, 
the scatter plots for 1980 and 1995 are dominated by a few outliers.
On the scatter plots each number represents a company; the link
with standard New Stock Exchange symbol names is as follows 
(full company names can be found on any
financial website such as for instance finance.yahoo.com).
\footnotesize
{\bf 1980:}
1=AA, 2=GE, 3=JNJ, 4=AXP, 5=GM, 6=KO, 
 7=T, 8=MMM, 9=UTX, 10=DD, 11=MO, 12=WMT, 13=MRK, 14=EK, 15=IP
{\bf 1985:}
 1=AA, 2=GE, 3=JNJ, 4=AXP, 5=JPM, 6=KO, 
 7=SBC, 8=C, 9=HON, 10=CAT, 11=MMM, 12=UTX, 13=DD, C14=MO, 15=WMT, 16=DIS,
 17=MRK, 18=EK, 19=IP
{\bf 1990:}
 1=AA, 2=KO, 3=INTC, 4=HON, 5=GE, 6=C, 
 7=JNJ, 8=MRK, 9=AXP, 10=SBC, C11=CAT,
 12=MMM, 13=UTX, 14=DD, 15=MO, 16=WMT, 17=DIS, 18=EK, 19=IP, 20=JPM
{\bf 1995:}
 1=AA, 2=GE, 3=JNJ, 4=AXP, 5=JPM, 6=KO, 
 7=SBC, 8=C, 9=HON, 10=CAT, 11=MMM,
 12=UTX, 13=DD, 14=MO, 15=WMT, 16=DIS, 17=INTC, 18=MRK, 19=EK, 20=IP.
\qL}
{\small \it Source: http://finance.yahoo.com}.
 \end{figure}
%% --------------------------------------------------
For 1980 the correlation is almost completely
due to Wal-Mart; moreover, the regression slope is much at variance
with the results for the other three years. Naturally, one would expect
that the slope of the regression becomes smaller when the initial date
which is selected approaches the date of the peak which is somewhere
between 2000 and 2002, and this is indeed the trend that we observe.
\qL

The significant correlation in cases 2 and 3 (see 
table 1) shows that the price multiplier rule applies to the
stocks in the DJ Industrial index whether one takes 1985 or 1990
as the initial year. For 1995 as the initial year, one observes 
the same effect, but there is little interest
in considering such a late starting point when previous ones work as well.
\qpar

We now turn to the companies in the DJ Transportation and Utility
indexes. The first index comprises 20 stocks, while the second contains
only 15. In each case a few companies need to be discarded because
the data are not available or because they did not pay dividends.
We are therefore left with samples which are fairly small; as a
result, the correlations although positive have broad confidence
intervals (see table 1).
In order to get a more significant result we have grouped together
all the stocks of the DJI, DJT and DJU; the corresponding graph is
shown in Fig.5. 
%%-----------------------------------------------
%%%% Fig.5
  \begin{figure}[tb]
    \centerline{\psfig{width=17cm,figure=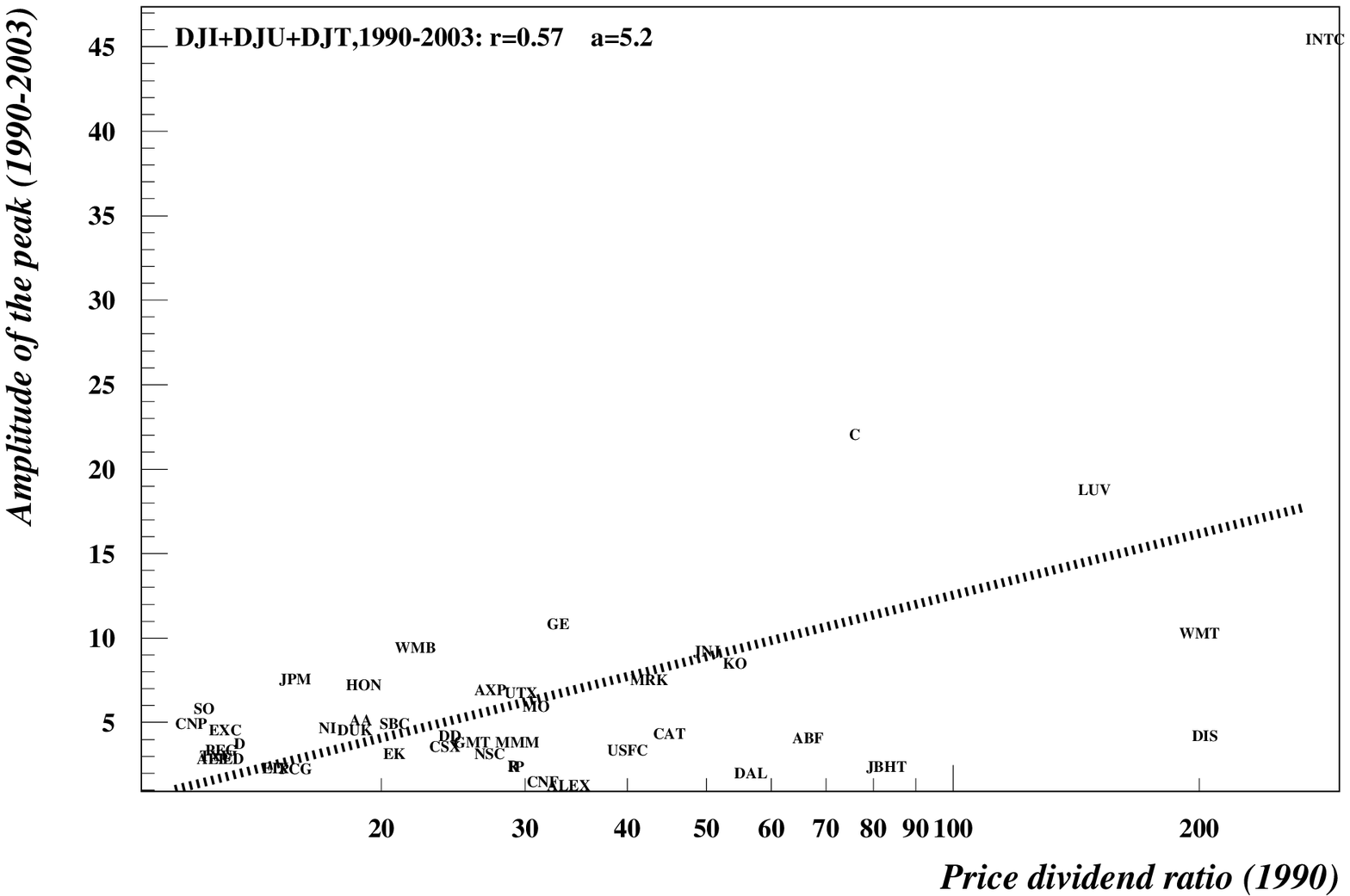}}
    {\bf Fig.5: Correlation between initial price-dividend
ratios and peak amplitudes.} 
{\small The stocks are those comprised in the DJ Industrial,
DJ Transportation and DJ Utility indexes  
for which price and dividend data were available on the Yahoo
website.}
{\small \it Source: http://finance.yahoo.com}.
 \end{figure}
%% --------------------------------------------------
On the one hand, the dispersion is increased because
the sample is less homogeneous, but on the other hand we now
have a sample comprising 44 stocks which results in reduced
confidence intervals for both the correlation and the regression
coefficients.
\qpar

In the next test we relaxed the homogeneity criterion and
considered a sample of companies drawn at random from the
set of 500 companies that make up the Standard
and Poor's index. In this case the correlation breaks down; table
1 shows that for 1990 at the starting year it is almost zero, whereas
for 1995 it is slightly positive but barely significant. It must be
said that what makes the matter worse is the fact that many 
of those companies 
in the information technology sector which experienced huge
price increases did not pay any dividend at all; in other words
their PDR was basically infinite and 
we cannot include these stocks into our sample.
In short, the cases which would provide the most
compelling evidence unfortunately
must be left out of the statistical analysis.

%%%%%%%%%%%%%%%%%%%%%%%%%%%%%%%

% TABLE 1

\begin{table}[htb]

 \small 

\vskip 3mm
\hrule
\vskip 0.5mm
\hrule
\vskip 2mm

\vskip 3mm
\centerline{{\bf Table 1\ Correlation between price-dividend ratio and
peak amplitude}: $ \hbox{Amplitude} = a\log(\hbox{PDR})+b $ } 
$$ \matrix{
\tvi 
\hbox{Case}& \hbox{Sample}\hfill & \hbox{Initial}  & \hbox{Number} 
& \hbox{Correlation} & \hbox{Confidence}  & a & b \cr
\tvi 
\hbox{number} & \hbox{}\hfill & \hbox{year}  & \hbox{of stocks}  & \hbox{} 
& \hbox{interval}  &  &  \cr 
\noalign{\hrule}
\qth 
1 &\hbox{DJI}\hfill & 1980 & 15 & 0.82 & 0.53,\ 0.94  & 107\pm 41 
&  -291 \pm 20 \cr
2 &\hbox{DJI}\hfill & 1985 & 19 & 0.46 & 0.003,\ 0.75  & 6.9\pm 6
& -10  \pm 4 \cr
3 &\hbox{DJI}\hfill & 1990 & 20 & 0.62 & 0.24,\ 0.83  &7.0 \pm 4
&  -17 \pm 3 \cr
4 &\hbox{DJI}\hfill & 1995 & 20 & 0.63  & 0.26,\ 0.84  & 2.8 \pm 1
&  -6.9 \pm 1 \cr
5 &\hbox{DJT}\hfill & 1990 & 11 & 0.69 & 0.15,\ 0.91  & 5.8 \pm 4
&  -18 \pm 2 \cr
6 &\hbox{DJU}\hfill & 1990 & 13 & 0.52 & -0.04,\ 0.83  & 5.3\pm 5
& -10  \pm 1 \cr
7 &\hbox{DJI+DJT+DJU}\hfill & 1990 & 44  & 0.58 & 0.34,\ 0.75  & 5.2\pm 2
&  -11 \pm 2 \cr
8 &\hbox{S\&P500}\hfill & 1990 & 60 & -0.03 & -0.28,\ 0.22  & 
&     \cr
9 &\hbox{S\&P500}\hfill & 1995 & 60 & 0.26 & 0.005,\ 0.48  &  
&   \cr
 & \hbox{}\hfill &  &  &  &    & 
&     \cr
 & \hbox{{\bf Average}}\hfill &  &  & \hbox{\bf 0.54} &   & \hbox{\bf 5.5}
&  \hbox{\bf 12} \cr
\qtb 
\hbox{}\hfill & &  &  &   &  &  \cr
\noalign{\hrule}
} $$

\vskip 1.5mm
Notes: One must keep in mind the fact
that because many of the high tech 
companies whose price shot up in the 1990s did not distribute any
dividend, they cannot be included in our samples;
this is of particular importance for the smaller companies considered in
the S\&P500 cases, but it also concerns a few major companies 
in the DJI sample (Microsoft in one example). 
In a few number of cases (IBM is one of the main examples)
dividends were duly distributed but for some reason are not recorded
in the Yahoo database.
When computing the average of the regression coefficients
$ a $ and $ b $ we discarded the first case which is obviously an
outlier.
\qL
{\it Source: http://finance.yahoo.com}
\vskip 2mm

\hrule
\vskip 0.5mm
\hrule

\normalsize

\end{table}

%% --------------------------------------------------------------

\qI{Conclusion}
Previous studies have shown that the strength of the
price multiplier effect that
is to say the level of the correlation
between initial prices and peak amplitudes 
can be seen
as a measure of the amount of speculative trading (as 
distinct from trades performed by users) which is going on in
a market. As on stock markets there are no users at all
(nobody buys a stock just to ``use it'');
one would expect 
the price multiplier effect to apply
to stocks even better than it applies to other assets.
However, for a number of reasons ranging from the high levels of 
noise and volatility to the dwindling importance
of dividends, 
the experimental test is far from being straightforward.
Nevertheless, when conducted in a proper way observation shows
that the price multiplier effect also holds for stocks provided
initial price-dividend ratios
are substituted for initial prices.
\qpar

At this point an observation is in order regarding the
dwindling role of dividends.
In recent times, companies more and more tended
to replace dividends by other gratifications among which 
buyback of shares is the most important. Usually companies buy back
their own shares at a price which is slightly higher than the 
market price; this
represents a one-time payment which can be seen
as a special dividend
replacing the payment of quarterly
dividends.
Over the past five or six years,
major companies such as Citigroup or IBM have devoted 
billions of dollars to buyback programs; it is estimated that
over this time interval
buybacks represented 1.3 trillion dollars that is to say
approximately 10 percent of the capitalization
of the New York Stock Exchange
in 2001 (source: Trimtabs.com cited in
WallstreetWishList.com). At the other end of the pipe,
new stocks are 
constantly created for instance through the distribution
of stock options, that is to say the right to buy a given
amount of stocks at a preset price and date.
Currently, the distribution of stock options is
limited mainly to employees,
but one could imagine stock holders to be rewarded in 
a similar way. That would have the obvious advantage of 
giving stock holders a strong incentive for pushing up
stock prices. 
\qL
At the present time it is difficult to say whether markets
will continue to move in this direction.
If they do
this would represent a significant watershed with 
respect to the way stock markets 
have been operating during the past two centuries.
With dividends losing much of their importance,
investors would increasingly have to rely on
capital gains. In other words, instead of being a cash
flow asset like commercial or residential property, stocks
would tend to
become an asset similar to rare coins, collector stamps
or antiquarian books. If it continues such an evolution
would significantly change the behavior of stock holders
during speculative episodes which was the topic that
we investigated in this paper.
\qpar

{\bf Acknowledgement} We express our gratitude to Professor
Bruce Mizrach for his kind assistance regarding data sets
for dividends.

\vskip 2cm

{\large \bf References}

\vskip 5mm

\qparr
[1] Roehner (B.M.) 2000: Speculative trading: the price
multiplier effect. The European Physical Journal B 14, 395-399.
Roehner (B.M.) 2001: Hidden collective factors in speculative
trading. Springer-Verlag. Berlin (chapter 6).

\qparr
[2] Roehner (B.M.) 2003: Patterns of speculation in real estate
and stocks. in Proceedings of the Nikkei Workshop in Econophysics
(Tokyo, November 2002). Springer-Verlag. Berling. in press.

\end{document}